\documentclass[a4paper]{article}

\usepackage{18lomcon}        
\usepackage{cite}             
\usepackage{graphicx}
\graphicspath{{img/}}

\newcommand{\cnb}{C$\nu$B}
\newcommand{\nob}{$N$-one-body}
\newcommand{\e}[1]{\ensuremath{\times 10^{#1}}}

\begin{document}


\title{NEUTRINO CLUSTERING IN THE MILKY WAY}

\author{Stefano Gariazzo \email{gariazzo.ific.uv.es} }

\affiliation{Instituto de F\'{\i}sica Corpuscular
(CSIC-Universitat de Val\`{e}ncia)\\ 
Parc Cient\'{\i}fic UV, C/ Catedr\'atico Jos\'e Beltr\'an, 2\\ E-46980 Paterna (Valencia), Spain}

\date{}
\maketitle


\begin{abstract}
The Cosmic Neutrino Background is a prediction
of the standard cosmological model, but it has been never observed directly.
In the experiments with the aim of detecting relic CNB neutrinos,
currently under development,
the expected event rate depends on the local density of relic neutrinos.
Since massive neutrinos can be attracted by the gravitational potential
of our galaxy and cluster around it,
a local overdensity of cosmic neutrinos should exist.
Considering the minimal masses guaranteed by neutrino oscillations,
we review the computation of the local density of relic neutrinos 
and we present realistic prospects for a PTOLEMY-like experiment.
\end{abstract}

\section{Introduction}

The standard cosmological model predicts the existence of a relic population
of neutrinos produced in the early Universe,
which is usually referred to as the Cosmic Neutrino Background (\cnb).
These neutrinos have nowadays a distribution
very close to a Fermi-Dirac
with an effective temperature of about 1.9~K, or $0.17$~meV.
This is small enough to say that at least two neutrinos over three
are non-relativistic today,
since neutrino oscillation experiments tell us
that the second lightest neutrino
must have a mass of at least $\sim8$~meV
\cite{deSalas:2017kay}.

Despite being the second most copious species in the Universe
after the photons of the Cosmic Microwave Background,
with a mean number density of $\sim330\,\rm{ cm}^{-3}$,
relic neutrinos are extremely difficult to detect, due to their very small energy.
The most interesting method that we can exploit for their direct detection
is the mechanism of neutrino capture (NC) in $\beta$-decaying nuclei,
acting through the process $\nu+n\rightarrow p+e^-$ \cite{Weinberg:1962zza}.
The interaction of a relic neutrino with a detector atom
forces the production of an electron that has an energy above
the endpoint of the standard $\beta$-decay by twice the neutrino mass:
this can be visible in the experiment if the energy resolution
is sufficient to distinguish the peak due to NC
from the standard $\beta$-decay events (see e.g.\ \cite{Long:2014zva}).

The PTOLEMY experiment \cite{Betts:2013uya},
currently under development, aims at detecting relic neutrinos
with a mass above $\sim$150~meV,
as the expected energy resolution of $\sim$100~meV allows.
The event rate from NC in the PTOLEMY detector,
built with 100~g of atomic tritium,
is expected to be of order ten per year,
if the mean number density of the \cnb\ is considered.

Even if neutrinos are very light, however,
they are expected to cluster around our galaxy,
thanks to the gravitational potential of the matter which forms it.
An increased local number density of relic neutrinos would correspond
to an increased event rate in the detector.
In the following we will show the results on the neutrino clustering in the Milky Way (MW),
published in ref.~\cite{deSalas:2017wtt},
which are based on the \nob\ simulation technique firstly presented in \cite{Ringwald:2004np},
and the corresponding prospects for the event rate in a PTOLEMY-like experiment.

\section{\nob\ simulations and the Milky Way}
We compute the clustering of relic neutrinos in the MW using the 
\nob\ technique~\cite{Ringwald:2004np},
which consists in independently evolving the trajectories of
a high number $N$ of neutrinos of mass $m_\nu$
in the gravitational potential of the MW,
from some early time until today,
sampling all the possible initial conditions
(neutrino position and momentum).
We assume initial homogeneity and spherical symmetry.
The final positions of these test particles are then employed
to reconstruct
the relic neutrino distribution in the MW today.
The ratio between the local number density at Earth, $n(m_\nu)$,
and the mean number density, $n_0$, gives the clustering factor
$f_c(m_\nu) = n(m_\nu) / n_0$,
which enters the calculation of the event rate (see next section).
In order to compute the neutrino trajectories,
we must adopt a description for the MW content and its time evolution.
We use results from the literature to describe the profiles
and the evolution of the MW content (dark matter, baryons)
as follows.

For the dark matter, we assume two possible descriptions for the halo:
the Navarro-Frenk-White (NFW) and the Einasto (EIN) profiles,
whose parametrizations are detailed in ref.~\cite{deSalas:2017wtt}.
The parameters which enter the NFW and EIN profiles are determined
using the astrophysical data from ref.~\cite{Pato:2015tja}
on the dark matter density.
The time dependence of the profiles is computed using
the standard evolution of the universe assuming the $\Lambda$CDM model,
the evolution of virial quantities and
the results of N-body simulations as given in ref.~\cite{Dutton:2014xda}.

The baryon content of the MW is described using the profiles
for the five components proposed in
ref.~\cite{Misiriotis:2006qq}:
stars, warm and cold dust, atomic and molecular hydrogen gas.
The time evolution of the total baryon profile is approximated
as a global renormalization constant,
which we obtain from N-body simulations
of MW-sized objects~\cite{Marinacci:2013mha}.
For more details, we refer to ref.~\cite{deSalas:2017wtt}.

\section{Clustering factors and PTOLEMY prospects}

\begin{figure}[t]
\centering
\includegraphics[width=0.49\textwidth]{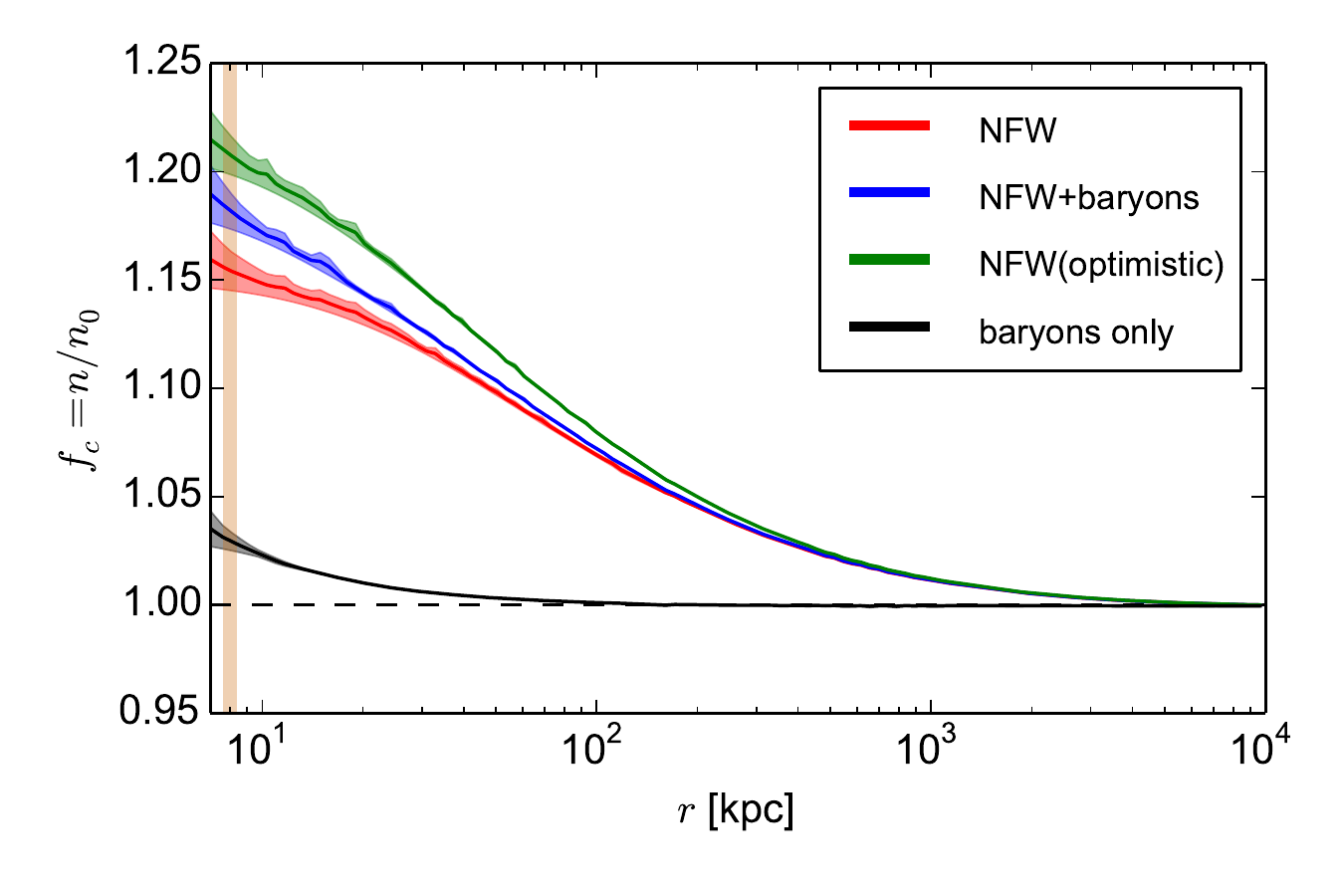}
\includegraphics[width=0.49\textwidth]{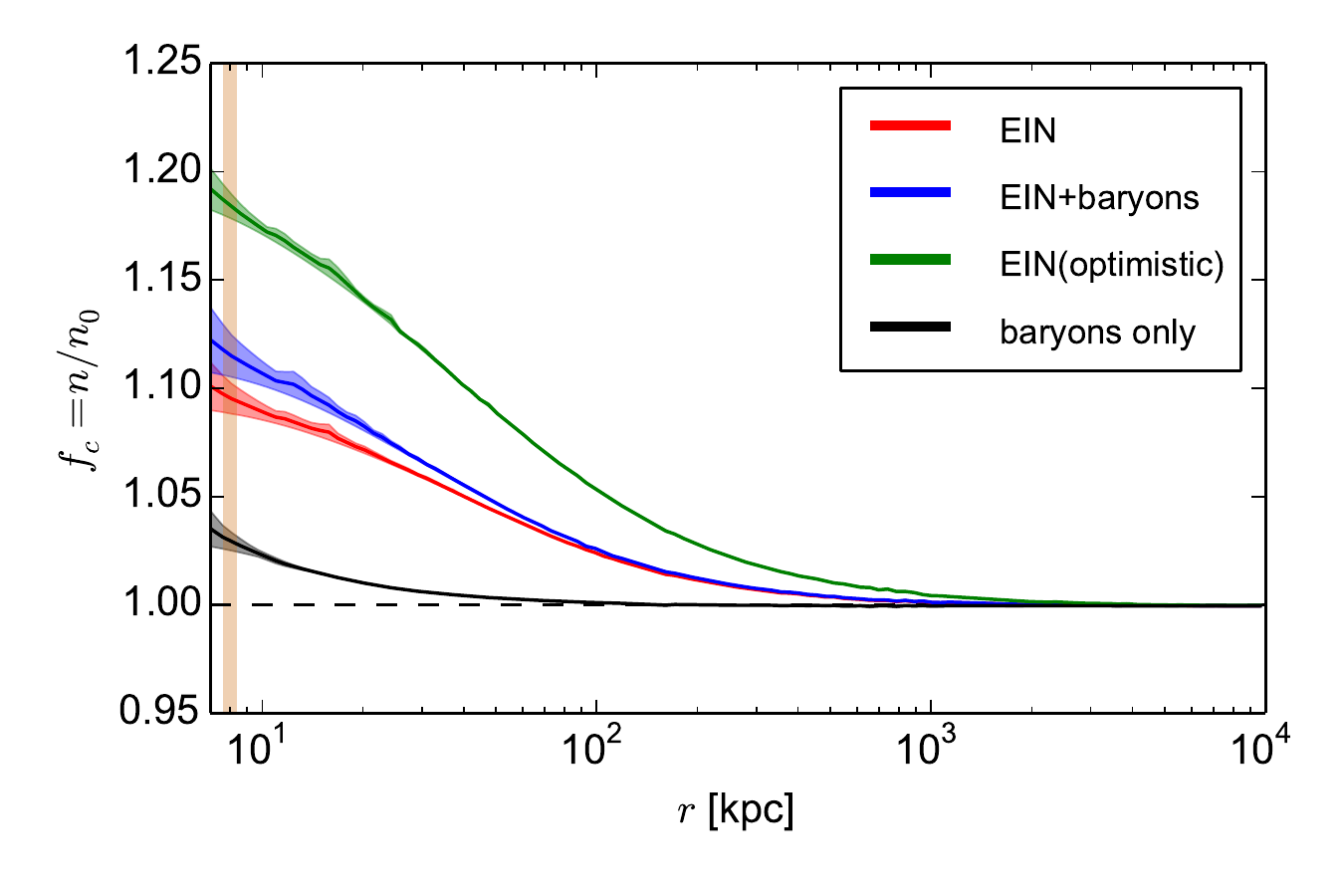}
\caption{\label{fig:clus60}
Profiles of the neutrino halo in the MW,
for a neutrino with a mass of 60~meV and
different parametrizations of the MW contents.
From ref.~\cite{deSalas:2017wtt}.}
\end{figure}

We firstly show the results obtained considering neutrinos
with nearly minimal masses.
Assuming that the heaviest mass eigenstate has a mass of $\sim$60~meV,
we run our \nob\ simulation and reconstruct
the profile of the neutrino halo
using different assumptions on the MW content,
as shown in figure~\ref{fig:clus60}, where
we plot the results obtained
using the NFW (EIN) dark matter profiles
in the left (right) panel,
alone and in combination with the MW baryons.
We can see comparing the two plots that the local neutrino density
can be up to 20\% larger than the mean neutrino density.
The most relevant source of error is represented by the MW structure,
since the results can significantly change
when different dark matter or baryon distributions are considered.

The rate of relic neutrino events expected in a PTOLEMY-like experiment
can be computed using \cite{Long:2014zva}:
\begin{equation}
 \label{eq:eventrate}
\Gamma_{\rm{C}\nu\rm{B}}^{}
  =
  \sum_{i=1}^{3}
  |U_{ei}|^2\,f_c(m_i)\,[n_{i,0}^{}(\nu_{h_R})+n_{i,0}^{}(\nu_{h_L})]\,N_T\,\bar\sigma\,,
\end{equation}
where $U_{ei}$ is the matrix element that encodes
the mixing between the neutrino mass eigenstate $i$ and the electron neutrino flavour,
$n_{i,0}^{}(\nu_{h_{R(L)}})$ is the mean number density
of right (left) helical neutrinos,
$N_T$ is the number of hydrogen nuclei in the detector and
$\bar\sigma\simeq3.834\e{-45}\,\rm{cm}^{-2}$.
Since it contains the mixing matrix elements,
eq.~(\ref{eq:eventrate}) tells us that
the event rate depends on the neutrino mass ordering,
when clustered neutrinos are considered.
In the case of normal mass ordering,
for which the mixing between the electron neutrino
and the heaviest mass eigenstate is the smallest,
the enhanced local neutrino density
has no impact on the expected event rate.
On the other hand, if the ordering is inverted,
the situation is opposite:
the $U_{e1}$ and $U_{e2}$ terms are large and the increase
in the event rate is directly proportional
to the clustering factor.

\begin{figure}[t]
\centering
\includegraphics[width=0.49\textwidth]{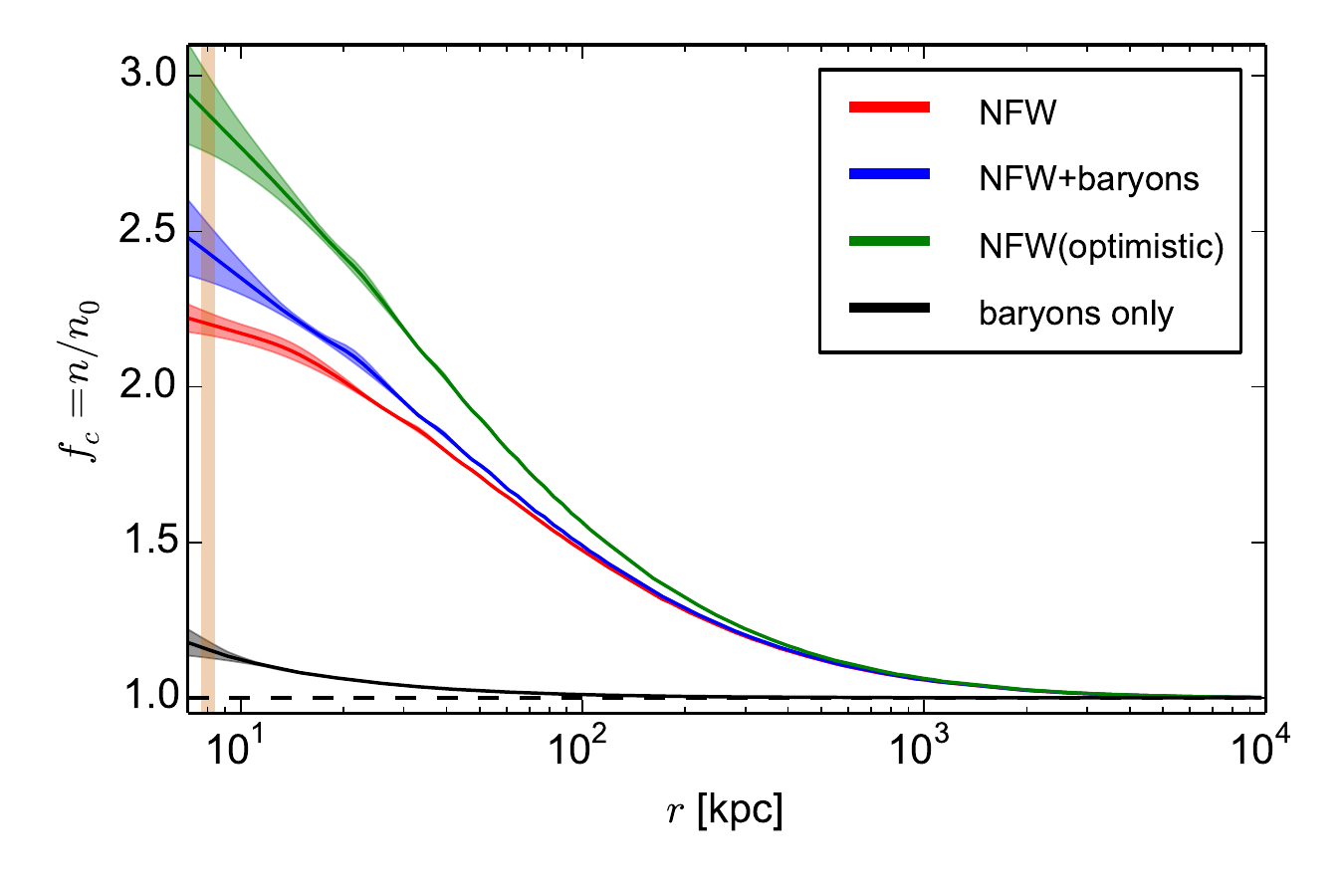}
\includegraphics[width=0.49\textwidth]{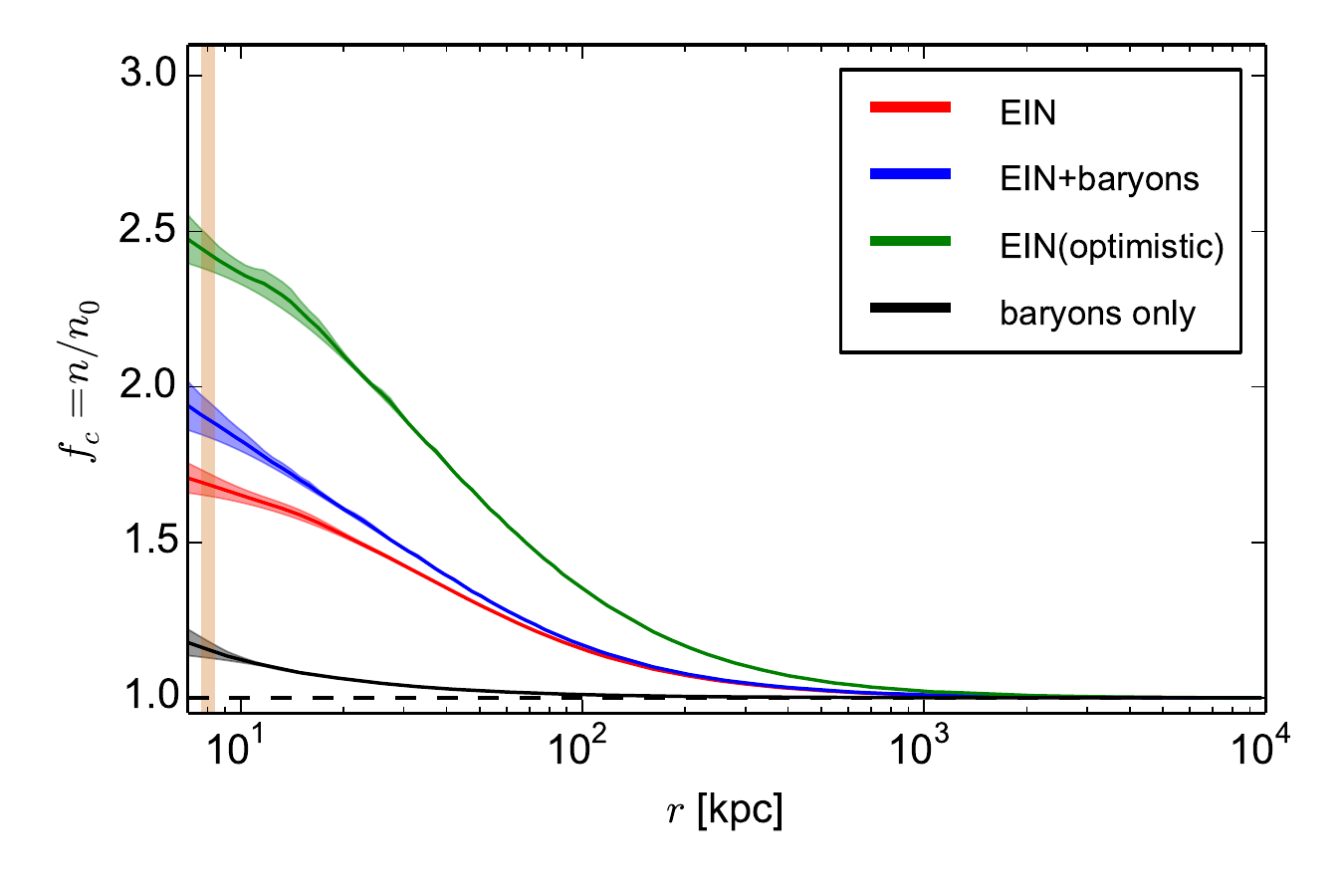}
\caption{\label{fig:clus150}
The same as figure~\ref{fig:clus60},
but for a neutrino mass of 150~meV.
From ref.~\cite{deSalas:2017wtt}.}
\end{figure}

The planned resolution for PTOLEMY,
unfortunately, will not allow a detection of 60~meV neutrinos.
For this reason we have also analysed
the case of neutrinos with a mass of 150~meV,
that should be the minimum mass detectable
by the experiment.
Considering this larger value of $m_\nu$,
neutrinos are practically degenerate in mass
and the event rate is not influenced by the mass ordering.
We get a clustering factor between 1.7 and 2.9,
as depicted in fig.~\ref{fig:clus150},
which corresponds to an increase of the event rate
by the same factor.
A precise determination of the event rate at PTOLEMY,
in this case,
would allow us to put constraints on the structure of our galaxy.

\section*{Acknowledgements}

Work supported by the Spanish grants
FPA2014-58183-P and
SEV-2014-0398 (MINECO),
and by the PROMETEOII/2014/084 (Generalitat Valenciana).

\section*{References}


\begin{thebibliography}{10}

\bibitem{deSalas:2017kay}
P.~F. de~Salas et~al.
\newblock {Status of neutrino oscillations 2017}.
\newblock Arxiv:1708.01186.
\newblock 2017.

\bibitem{Weinberg:1962zza}
S.~Weinberg.
\newblock {Universal Neutrino Degeneracy}.
\newblock {\em Phys. Rev.}, 128:1457--1473, 1962.

\bibitem{Long:2014zva}
A.~J. Long et~al.
\newblock {Detecting non-relativistic cosmic neutrinos by capture on tritium:
  phenomenology and physics potential}.
\newblock {\em JCAP}, 1408:038, 2014.

\bibitem{Betts:2013uya}
S.~Betts et~al.
\newblock {Development of a Relic Neutrino Detection Experiment at PTOLEMY:
  Princeton Tritium Observatory for Light, Early-Universe, Massive-Neutrino
  Yield}.
\newblock Arxiv:1307.4738.
\newblock 2013.

\bibitem{deSalas:2017wtt}
P.~F. de~Salas et~al.
\newblock {Calculation of the local density of relic neutrinos}.
\newblock {\em JCAP}, 09:034, 2017.

\bibitem{Ringwald:2004np}
A.~Ringwald and Y.~Y.~Y. Wong.
\newblock {Gravitational clustering of relic neutrinos and implications for
  their detection}.
\newblock {\em JCAP}, 0412:005, 2004.

\bibitem{Pato:2015tja}
M.~Pato and F.~Iocco.
\newblock {The Dark Matter Profile of the Milky Way: a Non-parametric
  Reconstruction}.
\newblock {\em Astrophys. J.}, 803(1):L3, 2015.

\bibitem{Dutton:2014xda}
A.~A. Dutton and A.~V. Macci\`{o}.
\newblock {Cold dark matter haloes in the Planck era: evolution of structural
  parameters for Einasto and NFW profiles}.
\newblock {\em Mon. Not. Roy. Astron. Soc.}, 441(4):3359--3374, 2014.

\bibitem{Misiriotis:2006qq}
A.~Misiriotis et~al.
\newblock {The distribution of the ISM in the Milky Way A three-dimensional
  large-scale model}.
\newblock {\em Astron. Astrophys.}, 459:113, 2006.

\bibitem{Marinacci:2013mha}
F.~Marinacci et~al.
\newblock {The formation of disc galaxies in high resolution moving-mesh
  cosmological simulations}.
\newblock {\em Mon. Not. Roy. Astron. Soc.}, 437(2):1750--1775, 2014.

\end{thebibliography}

\end{document}